\documentclass[preprint,12pt]{elsarticle}
\usepackage[utf8]{inputenc}
\usepackage{graphicx, url}
\usepackage{amssymb}
\usepackage{amsmath}
\usepackage{caption}
\usepackage{subcaption}
\usepackage{mathtools}
\usepackage[T1]{fontenc}
\usepackage{xcolor, soul}
\sethlcolor{yellow}

\journal{Advanced Powder Technology}

\begin{document}

\begin{frontmatter}

\title{Capturing the dynamics of a two orifice silo with the $\mu(I)$ model and extensions}

\author[massey]{Samuel K Irvine}
\author{Luke A Fullard}
\author[cant]{Daniel J Holland}
\author[vic]{Daniel A Clarke}
\author[massey]{Thomasin A Lynch \footnote{Corresponding author t.a.lynch@massey.ac.nz}}
\author[pyl]{Pierre-Yves Lagrée}

\affiliation[massey]{organization={School of Mathematical and Computational Sciences},
            addressline={Massey University}, 
            city={Palmerston North},
            country={New Zealand}}

\affiliation[cant]{organization={Department of Chemical and Process Engineering},
            addressline={University of Canterbury}, 
            city={Christchurch},
            country={New Zealand}}

\affiliation[vic]{organization={School of Chemical and Physical Sciences, Victoria University of Wellington},
            addressline={PO Box 600}, 
            city={Wellington},
            postcode={6140},
            country={New Zealand}}

\affiliation[pyl]{organization={Sorbonne Université, Institut Jean Le Rond d’Alembert},
            addressline={CNRS, UMR7190}, 
            city={Paris},
            postcode={75005},
            country={France}}

\begin{abstract}
Granular material in a silo with two openings can display a `flow rate dip', where a non-monotonic relationship between flow rate and orifice separation occurs. In this paper we study continuum modelling of the silo with two openings. We find that the $\mu(I)$ rheology can capture the flow rate dip if physically relevant friction parameters are used. We also extend the model by accounting for wall friction, dilatancy, and non-local effects. We find that accounting for the wall friction using a Hele-Shaw model better replicates the qualitative characteristics of the flow rate dip seen in experimental data, while dilatancy and non-local effects have very little effect on the qualitative characteristics of the mass flow rate dip. However, we find that all three of these factors have a significant impact on the mass flow rate, indicating that a continuum model which accurately predicts flow rate will need to account for these effects.
\end{abstract}



\begin{keyword}
Granular flow \sep Silo \sep non-local \sep dilatancy
\PACS 47.57.Gc
\MSC 74E20
\end{keyword}

\end{frontmatter}

\section{Introduction}
Whether it is in small-scale situations such as a salt shaker, or large industrial-scale situations such as blasted ore being mined from a draw-point, granular material is often stored in silos or silo-like domains. These domains can be challenging to model as they provide conditions for many different granular phenomena. In a flowing granular silo the flow behaviour can vary from the quasi-static regime where the material is static or nearly static, the dense regime where the granular material flows analogously to a fluid, and the dilute regime at the orifice where the material is in near free-fall. Developing models which can capture the flow of behaviour in such a complex domain is valuable to inform industrial silo design, as well as understanding granular flows in general.

One interesting flow rate phenomena is the flow rate `dip', which can arise when a silo has multiple orifices. Previous experiments done with spherical steel beads in a two opening silo have shown a monotonic decrease in flow rate as the orifice distance increases~\cite{xu2018inter}. However, experiments done using coarser, more industrially relevant materials result in a flow rate dip, where the flow rate for a silo with two openings in close proximity to each other is lower than the flow rate for a silo with larger separations between the openings~\cite{fullard2019dynamics}. A multiple orifice silo has been modelled using the kinematic and plasticity models~\cite{melo2007drawbody,melo2008kinematic}, however due to the flow rate being prescribed by the choice of parameters, the flow rate dip could not be analysed.

One method of modelling these flows is using Discrete Element modelling (DEM)~\cite{cundall1974computer}. This is a powerful method capable of predicting granular dynamics by considering interactions of pairs of particles, and can replicate some of the dynamics of the two orifice silo~\cite{zhang2016investigation}. However, because DEM requires modelling each particle individually it is computationally expensive, with the feasible number of particles that can be simulated being orders of magnitude smaller than the number of particles that are seen in an industrial context.

Alternatively, granular material may be modelled as a continuous pseudo-fluid. Such a continuum model could capture the desired macro-behaviour of granular flows while bypassing the computational overhead involved with modelling the micro-behaviour of granular material. As such a continuum model capable of accurately replicating the behaviour of granular material in a silo is relevant to many industries, however such a model is difficult to develop, with granular materials exhibiting many phenomena that are difficult to describe.

Some continuum models already exist, most notably the $\mu(I)$ model~\cite{gdr2004dense, jop2006constitutive}. This model captures the transition between quasi-static and dense flows (with dilute flows being predicted inaccurately~\cite{holyoake2012high}) using the dimensionless inertial number $I$. The inertial number represents a ratio between two timescales: how long it takes for granular material to move due to shear, and how long it takes for confining pressure to return dilated material to a resting state. As such, high $I$ represents the dilute regime where the material flows in a `gas-like' manner with shear rate being more important than confining pressure, while low $I$ represents the dense regime with a more `liquid-like' flow and longer lasting particle contacts, with the flow approaching a `solid-like' quasi-static regime as $I \to 0$. The inertial number is used to modify the frictional behaviour of the continuum model, which can be used to model flows in multiple different configurations.

However, while the $\mu(I)$ model can give good qualitative predictions for a silo~\cite{staron2012granular, staron2014continuum, zou2020discharge}, the quantitative flow rate predictions are not accurate~\cite{fullard2019quantifying}. The $\mu(I)$ rheology does not account for several key phenomena that can occur in a granular flow, which may explain this discrepancy. The $\mu(I)$ model applied to a pseudo-$2D$ silo does not account for the friction of the front and back walls, which other works have modelled as a Hele-Shaw like friction~\cite{zhou2017experiments}. Another effect which is not modelled is dilatancy, where a flowing mass of granular material will be less densely packed than a stationary mass~\cite{reynolds1885lvii}. This packing density likely significantly affects the mass flow rate for a silo, and as such will need to be accounted for in a continuum model which is expected to predict the mass flow rate. It also does not account for non-local effects~\cite{salvador2017modeling, henann2013predictive, kamrin2012nonlocal, bouzid2015non}, which are where the properties of flow at a point are determined by the flow behaviour of nearby material and not simply by the forces at that point. The $\mu(I)$ model is also not well behaved for all parameters and domains~\cite{barker2015well, heyman2017compressibility}.

Other continuum models exist and have been applied to silos. Several of these models, including plasticity models~\cite{irvine2016effect,nedderman1992statics}, the kinematic model~\cite{choi2005velocity,nedderman1979kinematic}, and the stochastic model~\cite{kamrin2008stochastic}, can give good descriptions for the qualitative behaviour of flow within the silo. However, each of these models have the flow rate determined by the choice of a parameter. This means that while these models can be useful for determining mixing behaviour and other such phenomena, they have little use when trying to determine quantitative flow rate behaviour.

\begin{figure}
\centering
\includegraphics[width=\textwidth]{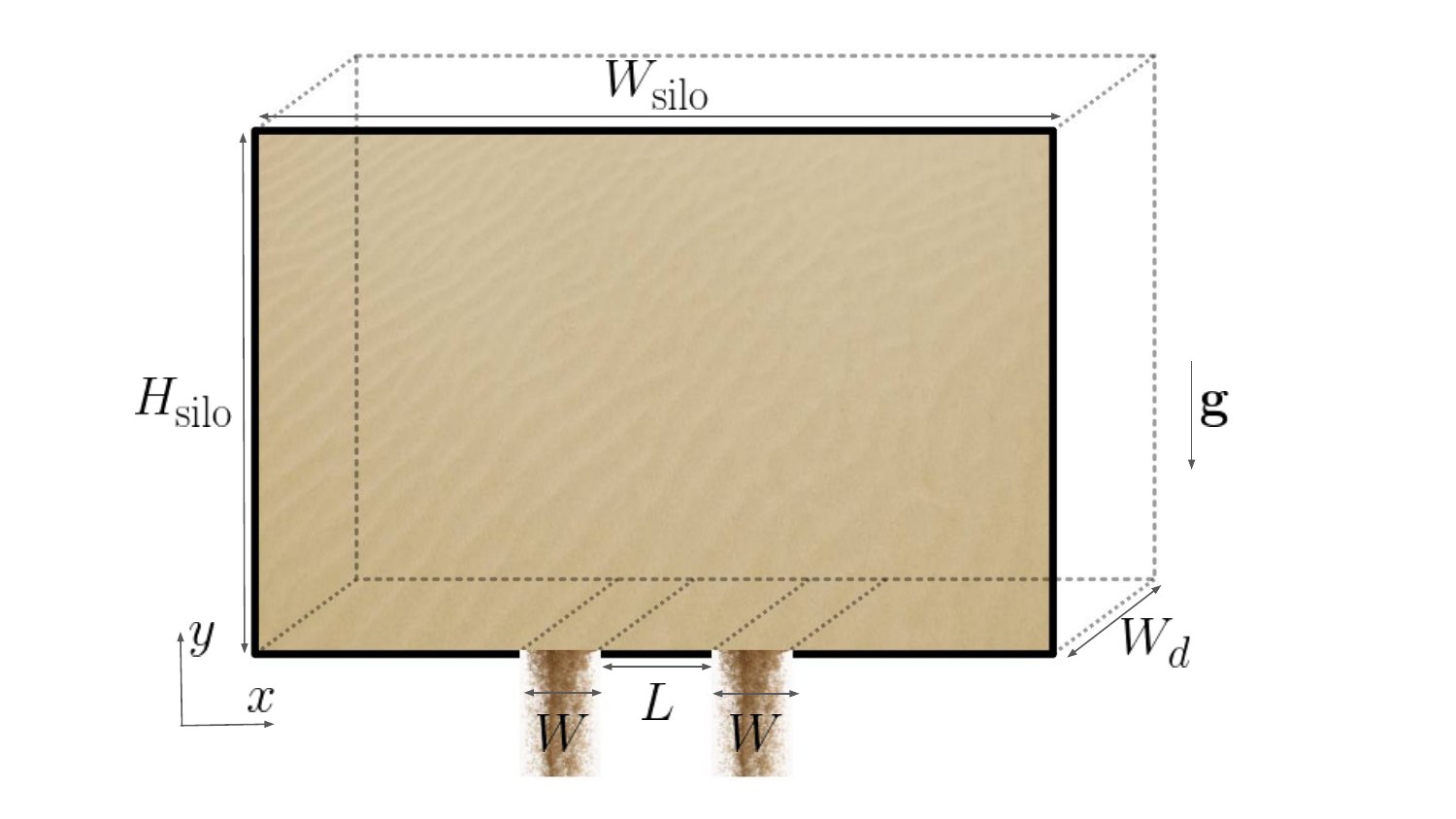}
\caption{Schematic of the system modelled. Two openings of diameter $W$ separated by a distance $L$ allow the granular material to drain. While the simulations are $2D$, experiments are done with some thickness $W_d$ with the assumption that $W_d << W_\textrm{silo}$ so that the silo can be considered $2D$ for the purpose of simulations, with $3D$ effects being accounted for in the Hele-Shaw extension.}
\label{fig:diagram}
\end{figure}

In this paper we examine the two opening silo, as depicted in Figure~\ref{fig:diagram} with parameters given in Table~\ref{tab:parameters}. Although this system is $3D$, we assume that the thickness is small so that it can be treated as a $2D$ system. We focus on using the $\mu(I)$ rheology, extending it to capture Hele-Shaw wall friction, dilatancy, and non-local effects. We additionally use the kinematic model for a simple comparison model for verification, while also demonstrating that it is unsuitable for double opening silos. We examine the flow rate phenomena using the $\mu(I)$ model, determining the effects each of the extensions have on mass flow rate magnitude and the flow rate dip. We use data from other experiments~\cite{fullard2019dynamics} for a comparison for the flow rate dip.

\section{Model and Implementation}
\subsection{The $\mu(I)$ model}

The $\mu(I)$ model, as described by~\cite{gdr2004dense} and extended to $3D$ by~\cite{jop2006constitutive}, is the baseline model that we use to describe the granular flows we are investigating. The model uses the incompressible Navier-Stokes Equations~(\ref{eq:Navier_Stokes})

\begin{equation}\label{eq:Navier_Stokes}
\begin{split} 
&\partial_t\mathbf{u}+\mathbf{u}\cdot\nabla\mathbf{u} = 
\frac{1}{\rho}\left[-\nabla p + \nabla\cdot(2\eta\mathbf{D})\right] -
\mathbf{g},\\
&\nabla\cdot\mathbf{u} = 0,
\end{split}
\end{equation}
where $\mathbf{u}$ is the velocity vector, $\rho=\phi\rho_\textrm{granular}+(1-\phi)\rho_\textrm{air}$ is the density derived from the packing fraction $\phi$ and the mix of material and gas density ($\rho_\textrm{granular}$ and $\rho_\textrm{air}$ respectively), $\mathbf{D}$ is the strain rate tensor given by
$\mathbf{D}=[\nabla\mathbf{u} + (\nabla\mathbf{u})^T]/2$, and $\mathbf{g}$ is the acceleration due to gravity. The relevant parameters are given in Table~\ref{tab:parameters}.

\begin{table}[t] 
\centering
\begin{tabular}{|l|l|l|l|l|}
\hline
Parameter & Value \\ \hline
Relative density of air $\rho_\textrm{air}/\rho_\textrm{granular}$ & $1.7\times10^{-3}$ \\ \hline
Orifice width $W$ & $9.375d$  \\ \hline
Domain width $W_\textrm{silo}$ & $150d$ \\ \hline
Domain height $H_\textrm{silo}$ & $150d$ \\ \hline
Separation length $L$ & $18.75d^*$ \\ \hline
Static friction $\mu_s$ & $0.62^*$ \\ \hline
Friction differential $\Delta\mu$ & $0.48^*$\\ \hline
Inertial number scaling $I_0$ & $0.6$ \\ \hline
Non-local model strength $A$ & $0.5^*$\\ \hline
Maximum solid fraction $\phi_\textrm{max}$ & $0.6$\\ \hline
Minimum solid fraction $\phi_\textrm{min}$ & $0.2$ \\ \hline
Solid fraction gradient $\phi_\textrm{grad}$ & $0.2^*$\\ \hline
Wall friction coefficient $F$ & $0.5^*$\\ \hline
\end{tabular}
\caption{Parameters used throughout this paper, with lengths being given as multiples of the particle diameter $d$. An asterisk indicates the value of the parameter varies in some simulations, with the value in this table indicating the value used when not otherwise specified.}
\label{tab:parameters}
\end{table}

These equations are combined with the $\mu(I)$ rheology, which assumes that the granular friction coefficient varies only on the inertial number $I$, which is given by
\begin{equation}
    I = \frac{|\dot{\gamma}|d}{\sqrt{p/\rho}},
\end{equation}
where $\dot{\gamma_{ij}}$ is the shear rate tensor given by
\begin{equation}
    \dot{\gamma_{ij}} = \frac{\partial u_i}{\partial x_j} + \frac{\partial u_j}{\partial x_i} = 2D_{ij},
\end{equation}
and $|\dot{\gamma}|$ is the second invariant of the shear rate tensor given by
\begin{equation}
    |\dot{\gamma}| = \sqrt{
       \frac{
            \dot{\gamma_{ij}}\dot{\gamma_{ij}}
        }{2}
    } = \sqrt{2D_{ij}D_{ij}},
\end{equation}
$d$ is the particle diameter, $p$ is the pressure, and $\rho$ is the material density.

While other formulations of the relationship of $\mu$ and $I$ are possible~\cite{kamrin2017hierarchy}, in this paper we make the assumption that the rheology is a kind of Coulomb friction with a coefficient of the form
\begin{equation} \label{eq:muI}
    \mu(I) = \mu_s + \frac{\Delta\mu}{(I/I_0+1)},
\end{equation}
where $\mu_s$, $\Delta\mu$, and $I_0$ are fitting parameters. Equation~\ref{eq:muI} defines the friction, which is implemented as an effective viscosity defined as
\begin{equation}\label{eq:effective_viscosity}
    \eta = \frac{\mu(I)p}{|\dot{\gamma}|},
\end{equation}
which is used in the Navier-Stokes equations. This gives us the basic $\mu(I)$ model.


In order to solve the incompressible Navier-Stokes equations from the $\mu(I)$ model we use the numerical scheme described by Popinet~\cite{popinet2003gerris}, which is implemented by Lagrée \& Staron~\cite{lagree2011granular} with the framework Basilisk~\cite{basilisk}. The Navier-Stokes Equations can be transformed using a projection method into a Poisson equation and a Helmholtz equation. The simulation uses a volume of fluid method representing the granular material as well as the air~\cite{lopez2015electrokinetic}. The boundary conditions are no-slip at the walls and base, and zero pressure at the top of the silo and inside the opening.

Another complication posed when implementing the $\mu(I)$ model is the static zones. A static region of material means that Equation~\ref{eq:effective_viscosity} gives an unbounded viscosity, and in a flat-bottomed silo there are some areas where there will be zero flow. To avoid divergent viscosity, a maximum viscosity $\eta_\textrm{max}$ is enforced, i.e. the finite $\eta^*$ is used, given by $\eta^* = min(\eta,\eta_\textrm{max})$, where $\eta$ is the viscosity calculated from Equation~\ref{eq:effective_viscosity} and $\eta_\textrm{max}$ is a large constant. This leads to a small creeping flow in these regions which should be static in a physical silo. For our simulations we use $\eta_\textrm{max} = 1800 \sqrt{\rho^2d^5g^{-1}}$, which is sufficiently large such that the creeping flow is negligible compared to the flow from the silo draining due to gravity.

It should be noted that this model assumes that the stress and strain rate tensors are aligned. The $\mu(I)$ model relies on visco-plastic theory which assumes that these tensors are in the same direction, which is not always the case~\cite{cortet2009relevance}. Also, the model is not well-posed, so may fail for some parameters and some domains~\cite{barker2015well, heyman2017compressibility}. The parameters and domain studied in this paper do not display evidence of numerical instabilities and the inertial number is in the well-posed region outside of the `static' zones~\cite{barker2015well}, however caution should be applied when using these models.

In order to avoid applying different boundary conditions within the length of a single cell, all lengths are chosen to be some integer multiple of the cell width (which due to the tree-based discretization implementation used~\cite{van2018towards} is the domain width divided by $2^n$, where $n$ determines the resolution of the simulation and is chosen to be $n=8$ for this work).

\subsubsection{Hele-Shaw friction}

In implementing the $\mu(I)$ model in a $2D$ silo, we assume that the granular material is in a true two dimensional system. In reality, the experiments are done in a system that only approximates two dimensions, with front and back walls being separated by a small non-zero distance~\cite{fullard2019dynamics}. It is possible to improve the comparison to experimental data by accounting for the friction between the granular material and the front and back walls~\cite{zhou2016ejection}.

The wall friction provides a force $\mu_w p$ in the opposite direction of the $2D$ flow. In order to account for this wall friction, we modify Equation~\ref{eq:Navier_Stokes}. Since we assume that the variation of flow over the thickness is negligible (i.e. the flow is only displaying $2D$-like behaviour), the velocity can be updated by calculating the Hele-Shaw friction, which results in an extra term in the Navier-Stokes equations
\begin{equation}\label{eq:wall_NS}
\begin{split} 
&\partial_t\mathbf{u}+\mathbf{u}\cdot\nabla\mathbf{u} = 
\frac{1}{\rho}\left[-\nabla p + \nabla\cdot(2\eta\mathbf{D})\right] -
\mathbf{g}-\frac{2\mu_w  p \mathbf{u}}{W_{d}|\mathbf{u}|},\\
&\nabla\cdot\mathbf{u} = 0,
\end{split}
\end{equation}
where $\mu_w$ is the friction coefficient between the granular material and the walls, $W_{d}$ is the distance between the front and back wall, $p$ is the pressure, and $u$ is the velocity.

To implement wall friction, we calculate the Navier-Stokes equations, then apply an update to the velocity from the front and back wall friction~\cite{zhou2017experiments}. We take the velocity already calculated, $u$, and calculate an update to the velocity term
\begin{equation}\label{eq:wall}
    \Delta \mathbf{u} = -\frac{2\mu_w  p \mathbf{u} \Delta t}{W_{d}|\mathbf{u}|},
\end{equation}
with $\Delta t$ as the time step. If the update to the velocity is greater than the friction-free velocity (which could result in unphysical upward flows), the velocity is instead set to zero. This corresponds to friction being greater than the other net forces and completely arresting the flow. This updated velocity is then fed into the unmodified Navier-Stokes equations for the next time step.

Note that the parameters $\mu_w$ and $W_d$ are both constant and are not used elsewhere. Therefore, in order to simplify the parameter set we combine them by choosing a parameter $F = \frac{2\mu_w}{W_d}$ which we vary to test the wall friction model. Then Equation~\ref{eq:wall} becomes
\begin{equation}
   \Delta \mathbf{u} = -F p \Delta t\frac{\mathbf{u}}{|\mathbf{u}|}. \label{eq:hele-shaw}
\end{equation}
Using the amaranth experimental data as a comparison~\cite{fullard2019dynamics}, the $F$ value is $4\mu_w$. It is unclear what value of $\mu_w$ should be used, with values from $0.1$ to $0.25$ being used in similar domains~\cite{shafaei2016analytical, zhou2017experiments}, corresponding to $F=0.4$ and $F=1$ respectively. The complications of dynamic and rolling friction and the influence of wall smoothness make determining the correct value difficult. We find the simulations become unstable for $F>0.5$, so we focus on cases where $F\leq0.5$ which is sufficient to show that wall friction has a significant impact.

\subsubsection{Dilatancy}

While assuming the granular material is incompressible is a first order approximation that simplifies the numerical methods, in reality granular material is known to dilate when sheared~\cite{andreotti2013granular, reynolds1885lvii}. Dilatancy introduces compressibility, meaning that models which fully take into account the compressible Navier-Stokes equations are complex~\cite{barker2017well, pailha2009two, bouchut2021dilatancy}. However, over the range of inertial numbers relevant to a silo the packing fraction seems to linearly decrease with the inertial number~$I$~\cite{hurley2015friction}. We can implement this linear dependence as a simple model for dilatancy, with 
\begin{equation} \label{eq:linear_dilatancy}
    \phi = \textrm{max}(\phi_\textrm{max}-\phi_gI, \phi_\textrm{min})
\end{equation}
where $\phi_g$ is the linear gradient parameter, $\phi_\textrm{max}$ is a constant representing the packing fraction for a material that is not being sheared ($0.6$ is used throughout this paper), and the minimum packing fraction is set to $\phi_\textrm{min}=0.2$, in order to prevent non-physical negative packing fractions. While other formations of the relationship of $\phi$ to $I$ are sometimes used~\cite{robinson2021evidence}, for simplicity we limit our scope to only consider linear dependence. This packing fraction is used to calculate the bulk density of the material. With non-homogeneous density the incompressible assumption $\nabla\cdot u = 0$ is replaced with a `source term', which is derived from conservation of mass. This source term is given by
\begin{equation}
    \nabla \cdot \mathbf{u} = \frac{1}{\rho}\left ( \frac{\partial\rho}{\partial t} + \mathbf{u}\cdot\nabla\rho \right ),
\end{equation}
where $\rho$ is the bulk density. The dilatancy then has flow on effects to the rest of the relevant simulation fields, and is incorporated into the mass flow rate (which is directly proportional to the volumetric flow rate in the absence of dilatancy). The bulk density is dependent both on the material density as well as the packing fraction of the granular material, determined by the linear dilatancy model in Equation~\ref{eq:linear_dilatancy} (if there is no relationship between $\phi$ and $I$, the incompressible assumption is recovered). The source term dilatancy model approximates the source term and incorporates it in the projection method.

\subsubsection{Non-local effects}

The $\mu(I)$ model and extensions are all local models, meaning that the properties of flow at a point are determined only by other properties at that point. However, granular material exhibits non-local effects, meaning that the behaviour of the material around a given point can affect the flow at that point. An example of a geometry which displays non-local effects clearly is flow down a slope, where there is a difference between the angle at which flow stops when the slope is lowered and the angle at which it starts when the slope is raised~\cite{pouliquen2009non}. This difference exists because the flowing particles agitate their neighbours, maintaining flow when a local model would predict no flow is possible. Another example is an annular shear cell, where local models predict flow sharply going to zero in the areas where the yield criterion is not met, while in experiments an exponential decay is observed~\cite{henann2013predictive}.

In order to capture these non-local effects, we use a granular fluidity model~\cite{kamrin2012nonlocal, salvador2017modeling, faroux2021coupling}. This model finds the local `fluidity' (which can be conceptualised as an inverse viscosity), and then `spreads out' the fluidity into nearby regions, representing the agitation caused by flowing particles in a neighbourhood that create non-local effects. The fluidity $g$ (not to be confused with gravity) is related to the $\mu$ value by the relation $|\dot{\gamma}| = \mu g$. The fluidity has some local value $g_l$, which corresponds to the fluidity if there were no non-local effects. The local $g_l$ is then `spread out' by the Laplacian term
\begin{equation}
    g = g_l + \xi(\mu(I))^2 \nabla^2 g,
\end{equation}
where $\xi(\mu(I))$ is given by
\begin{equation}
    \xi(\mu(I)) = A \sqrt{\frac{\mu_s + \Delta\mu - \mu(I)}{\Delta\mu(\mu(I)-\mu_s)}} \label{eq:non_local}
\end{equation}
where $A$ is a parameter which determines how strong the non-local effects are, with $A = 0$ corresponding to a purely local model. With the fluidity $g$ calculated, the relations $\dot{\gamma} = \mu g$ and $\eta = \frac{\mu P}{\dot{\gamma}}$ can be combined to give the effective viscosity as
\begin{equation}
    \eta = \frac{p}{g},
\end{equation}
which is used in Equation~\ref{eq:Navier_Stokes} in the same manner as the base $\mu(I)$ model. Boundary conditions are zero fluidity $g=0$ at the walls (corresponding to infinite viscosity i.e. no-slip) and zero normal flux $g_n = 0$ at the opening and top. A similar non-local fluidity model makes the model well-posed~\cite{kamrin2019non} so it is possible that this model is well-posed, however additional research is required. In addition, combining non-local fluidity with the additional models for Hele-Shaw and dilatancy has unknown stability.

\subsection{Kinematic model}
The Nedderman T{\"u}z{\"u}n kinematic model~\cite{nedderman1979kinematic} makes the assumption that a gradient in vertical velocity causes horizontal motion, i.e. $u = -B\frac{\partial v}{\partial x}$, where $u$ and $v$ are the horizontal and vertical velocity respectively and $B$ is a constant. This assumption, combined with the assumption that material is incompressible, gives a relatively simple model of granular flow taking the form of the heat equation~\cite{melo2008kinematic},
\begin{equation}
\frac{\partial v}{\partial y} = B\frac{\partial^2 v}{\partial x^2}.
\end{equation}
This partial differential equation can be solved exactly across the infinite half-plane with a Dirac delta $v=\delta(x)$ boundary condition at $y=0$, corresponding to an infinitesimal orifice. The solution of this boundary value problem is
\begin{equation}\label{eq:kinematic1hole}
v = \frac{-Q}{\sqrt{4\pi B y}}\textrm{exp} \left( -\frac{x^2}{4By} \right),
\end{equation}
where $Q$ is the flow rate. For a silo with two openings separated by some distance $L$, the boundary condition at $y=0$ is $v = \delta(x-L/2) + \delta(x+L/2)$, resulting in a solution of the form
\begin{equation}\label{eq:kinematic2hole}
v = \frac{-Q}{\sqrt{4\pi B y}} \left[ \textrm{exp} \left( \frac{(x-L/2)^2}{-4By} \right) + \textrm{exp} \left( \frac{(x+L/2)^2}{-4By} \right) \right].
\end{equation}
Note that the kinematic model is scaled by the prescribed flow rate, $Q$. For the two opening case the flow rate for any non-zero separation is simply twice the flow rate of a single orifice (i.e. $2Q$) and does not vary with separation. In order to account for orifice interaction a Beverloo-like relation of the form $Q(L)$ would need to be developed. As such, in its original form, the kinematic model of granular flow from a silo is unable to capture either the monotonic decrease seen in previous experiments \cite{zhang2016investigation,xu2018inter} or the flow-rate dip phenomenon. As such, we limit the use of the kinematic model to only model the single orifice silo.

\section{Results}

\subsection{Single orifice silo} \label{sec:single_results}

\begin{figure}
     \centering
     \begin{subfigure}[t]{0.47\textwidth}
         \centering
         \includegraphics[width=\textwidth]{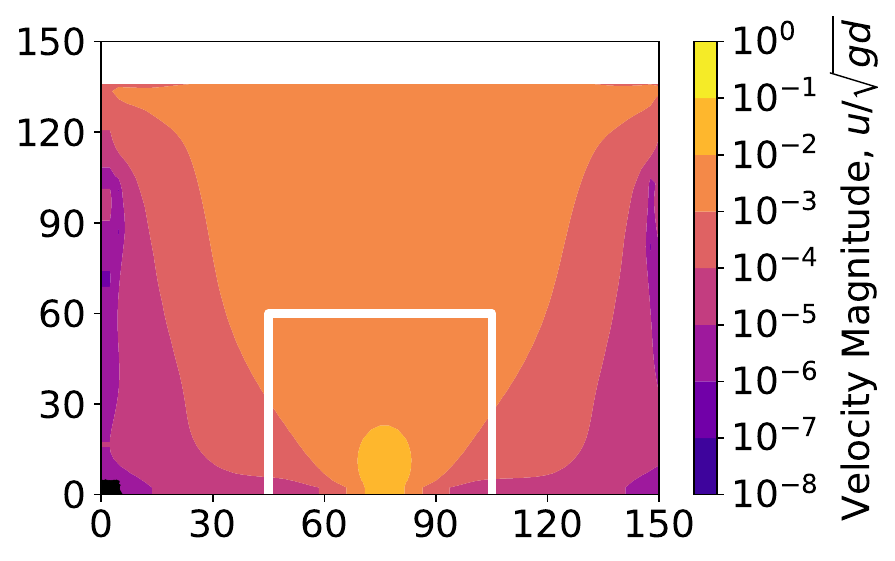}
     \end{subfigure}
     \hfill
     \begin{subfigure}[t]{0.47\textwidth}
         \centering
         \includegraphics[width=\textwidth]{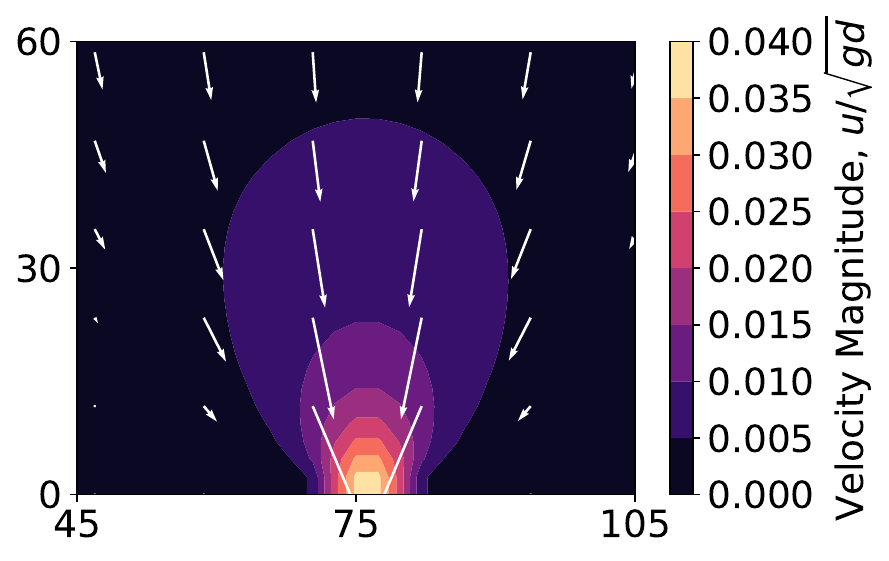}
     \end{subfigure}
        \caption{Velocity magnitude for a single opening silo, with a log scale (left) and linear scale zoomed to the opening (right). The box in the log plot indicates the area covered in the linear plot. The $\mu(I)$ parameters used are ($\mu_s$, $\Delta\mu$, $I_0$) = ($0.62$, $0.48$, $0.6$), with an opening diameter of $0.625$.}
        \label{fig:contour}
\end{figure}

We test the $\mu(I)$ model by examining the single orifice silo, which is a well studied $2D$ adaption of a common industrial domain. By applying the base $\mu(I)$ model to the single orifice silo, we obtain the velocity profile shown in Figure~\ref{fig:contour}. The flow matches the qualitative behaviour we expect to see in a flat-bottom silo, with static zones in the corners (with near-static creeping flow caused by the regularisation on viscosity), high flow near the orifice decreasing rapidly as points further from the orifice are considered, and `mass flow' like behavior in the upper regions of the silo.

We also use the Nedderman T{\"u}z{\"u}n kinematic model~\cite{nedderman1979kinematic} for comparison, which finds the vertical velocity $v$ to be
\begin{equation}
    v = \frac{-Q}{\sqrt{4\pi By}}\textrm{exp}\left(\frac{-x^2}{4By}\right),
\end{equation}
where $Q$ is the mass flow rate (which is prescribed), $B$ is a fitting parameter, and $x, y$ are the horizontal and vertical distance from the orifice, which is modelled as a single point. Note that even if the kinematic model gives a sufficiently accurate prediction for the qualitative behaviour, a limitation of the model is that the flow rate is prescribed by $Q$ and so provides no prediction for flow rate. As such, we use the kinematic model for comparing qualitative behaviour to the $\mu(I)$ model before using the $\mu(I)$ model to examine the flow rate behaviours.

\begin{figure}
\centering
\includegraphics[width=0.47\textwidth]{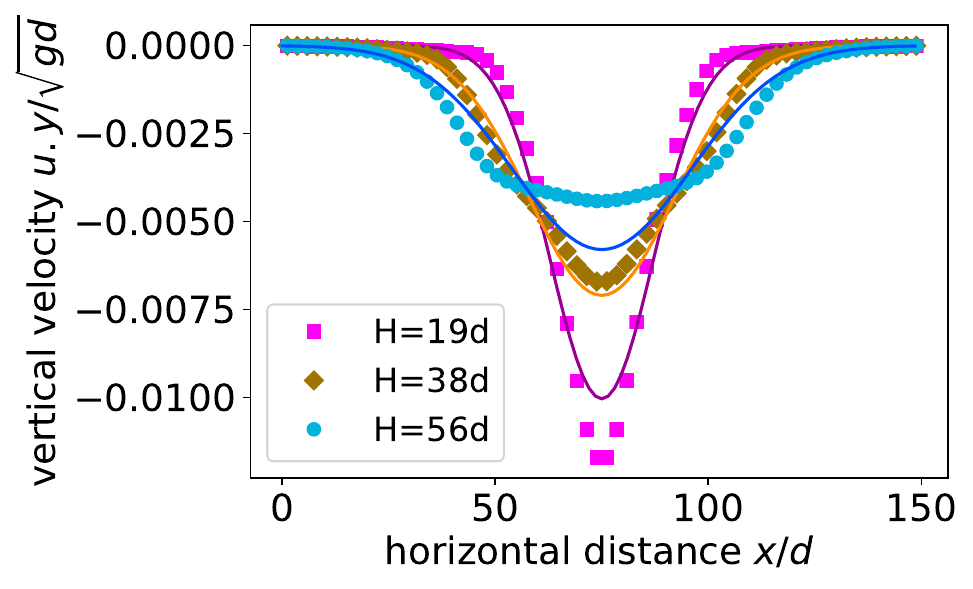}
\caption{Kinematic model (lines) compared with $\mu(I)$ model (points) in a single opening $2D$ silo. The vertical velocity from simulations and kinematic model are compared for various different heights. One set of kinematic parameters is fitted over the entire silo.}
\label{fig:kinematic_compare}
\end{figure}

A comparison between the kinematic model and the $\mu(I)$ model is shown in Figure~\ref{fig:kinematic_compare}, which shows the vertical velocity at various different horizontal slices in the silo. The $\mu(I)$ model predicts a `flatter' velocity curve for large heights than the kinematic model. We expect that in the upper middle of the silo the flow transitions to plug flow, which the kinematic model is unable to capture. This indicates that plug flow is a phenomena that the $\mu(I)$ model can capture that the kinematic model cannot. While the kinematic model could give a closer fit for any particular height if the $B$ parameter is varied, this is requires a height dependent $B$ parameter which is not justified by the assumptions of the kinematic model.

To further validate the $\mu(I)$ model we examine the effect of orifice width on the mass flow rate. For a silo with a single orifice we expect that the flow rate follows the Beverloo-Hagen relation, which for a $2D$ silo has the form~\cite{zhou2017experiments}
\begin{equation}
    Q = C\rho\sqrt{g}(W-kd)^{1.5},
\end{equation}
where $Q$ is the $2D$ mass flow rate, $C$ and $k$ are fitting parameters, $\rho$ is the bulk density, $g$ is here gravitational acceleration, $W$ is the orifice width, and $d$ is the particle diameter. Experiments suggest the $k$ should be in the range $1<k<2$, although theoretical arguments have been made that k should be $1$ exactly~\cite{beverloo1961flow,mankoc2007flow}.

\begin{figure}
     \centering
     \begin{subfigure}[t]{0.47\textwidth}
         \centering
         \includegraphics[width=\textwidth]{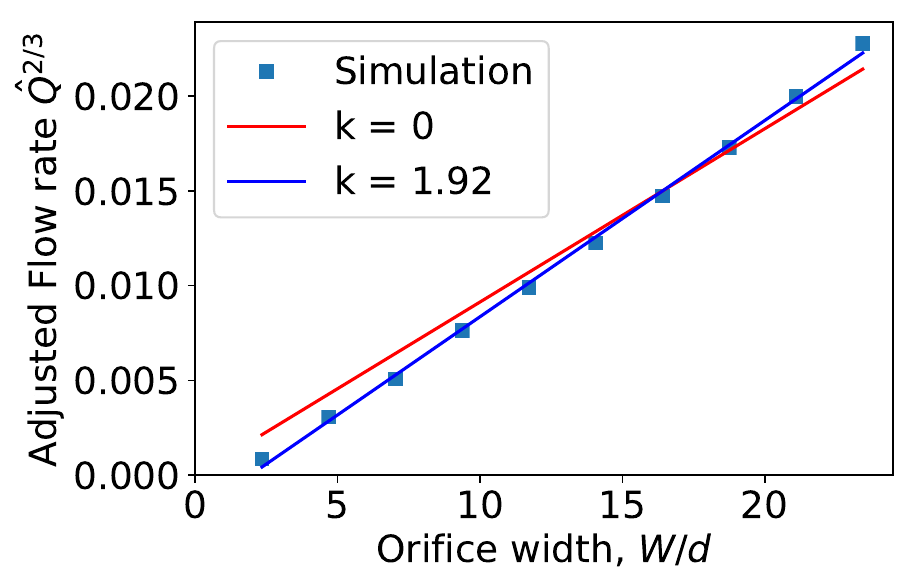}
     \end{subfigure}
     \hfill
     \begin{subfigure}[t]{0.47\textwidth}
         \centering
         \includegraphics[width=\textwidth]{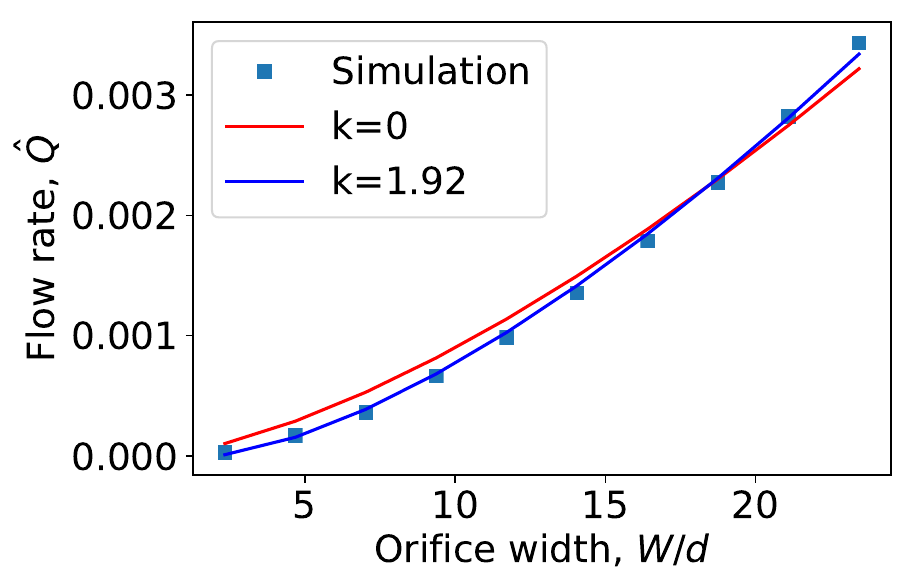}
     \end{subfigure}
        \caption{Flow rate $\hat{Q}=Q/\sqrt{gd^5}$ compared to orifice width on a $\frac{2}{3}$ flow rate scale (left) and a linear scale (right), with $k=0$ and fitted $k=1.92$}
        \label{fig:beverloo}
\end{figure}

In Figure~\ref{fig:beverloo} we see a comparison of the Beverloo relation with the simulated $\mu(I)$ silo. The flow rate is measured by taking the mass of granular material remaining over time and fitting a straight line (omitting the first time unit of flow to avoid transitional effects from the initial conditions). For the $k=0$ case we see results consistent with the expected $Q \propto W^{1.5}$ relationship. The $kd$ term is often attributed to single particle interactions, so we do not expect that it would be captured by this continuum model. Nonetheless, the $\mu(I)$ model seems to capture this $kd$ term with $k\approx1.92$, and the Beverloo model has a clear discrepancy with the simulations when this term is omitted. This may suggest that the $kd$ term is not caused by individual particle interactions or the ``empty annulus" theory, but is instead primarily caused by the friction near the orifice. Note that the low orifice widths represent unphysical silos; for orifice widths less than $~5d$, jamming becomes a significant factor and no meaningful flow rate measurement can be achieved experimentally. In contrast, the $\mu(I)$ model gives non-zero predictions for the flow rate for all silos where the orifice width is greater than zero.

\begin{figure}
     \centering
     \begin{subfigure}[t]{0.47\textwidth}
         \centering
         \includegraphics[width=\textwidth]{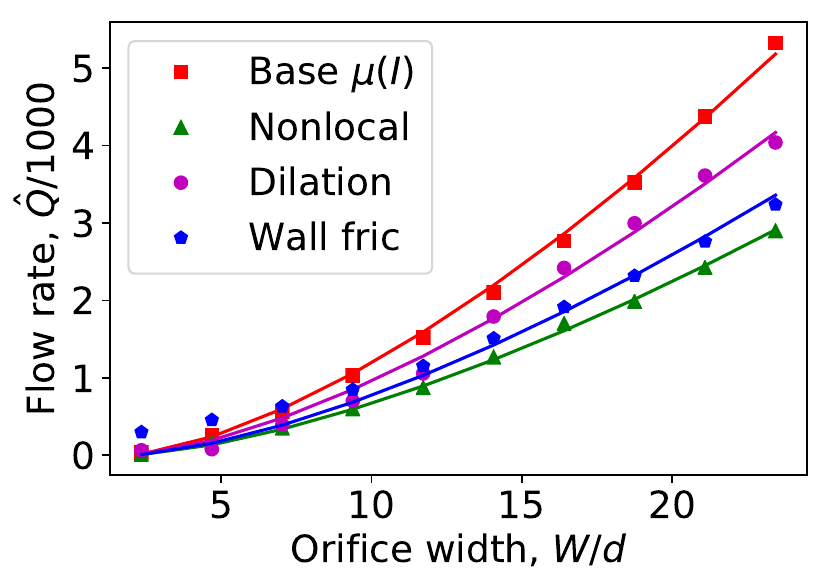}
     \end{subfigure}
     \hfill
     \begin{subfigure}[t]{0.47\textwidth}
         \centering
         \includegraphics[width=\textwidth]{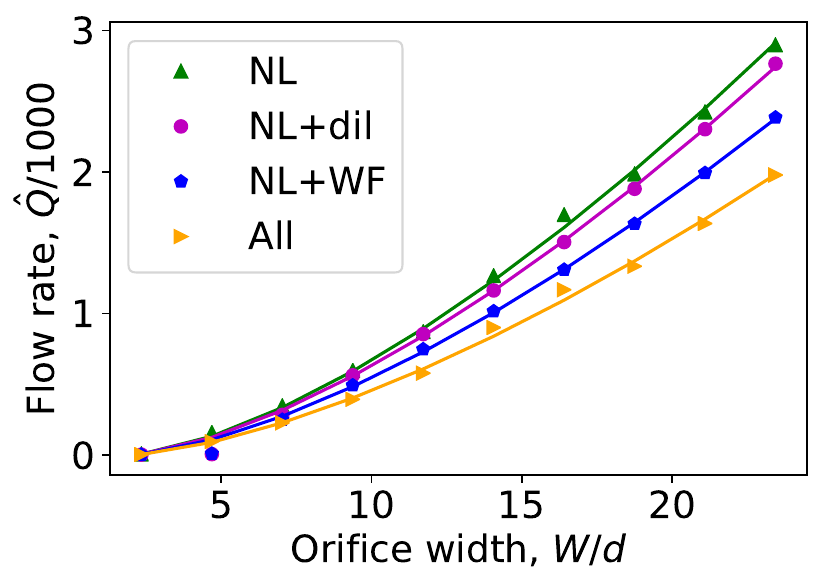}
     \end{subfigure}
        \caption{Mass flow rate $\hat{Q}=Q/\sqrt{gd^5}$ for various single orifice widths and combinations of non-local effects, dilatancy, and wall friction. The points show the simulation, while the lines give the fitted Beverloo relation with $k=1.92$. When relevant, the non-local strength is set to $A=0.5$, dilatancy gradient is set to $\phi_g=0.2$, and wall friction strength is set to $F=0.5$.}
        \label{fig:beverloo_additions}
\end{figure}

In Figure~\ref{fig:beverloo_additions}, the Beverloo relation is examined when non-local effects, dilatancy, and/or wall friction are considered. Non-local effects and dilatancy seem to decrease the flow rate, with non-local effects having the stronger effect for these parameter values. Wall friction also seems to increase the flow rate for high orifice widths, however low orifice widths it seems to increase the flow rate. This could be captured by decreasing the $kd$ term, however this results in lower $kd$ values seen in some experiments and theories. However, it should again be stressed that this analysis is done including non-physical silos with orifice widths too small to give consistent flow, and as such this analysis may have limited application to physical silos.

In Figure~\ref{fig:beverloo_additions} we also combine non-local effects with combinations of dilatancy and wall friction. Adding dilatancy or wall friction when non-local effects are present decreases the flow rate, but to a lesser extent compared to when non-local effects are absent. The increased flow rate for low orifice widths also seems to be suppressed when non-local effects are present. Any continuum model wishing to capture the flow rate accurately will need to account for each of these effects.

\subsection{Double orifice flow rate dip}

\begin{figure}
\centering
\includegraphics[width=0.47\textwidth]{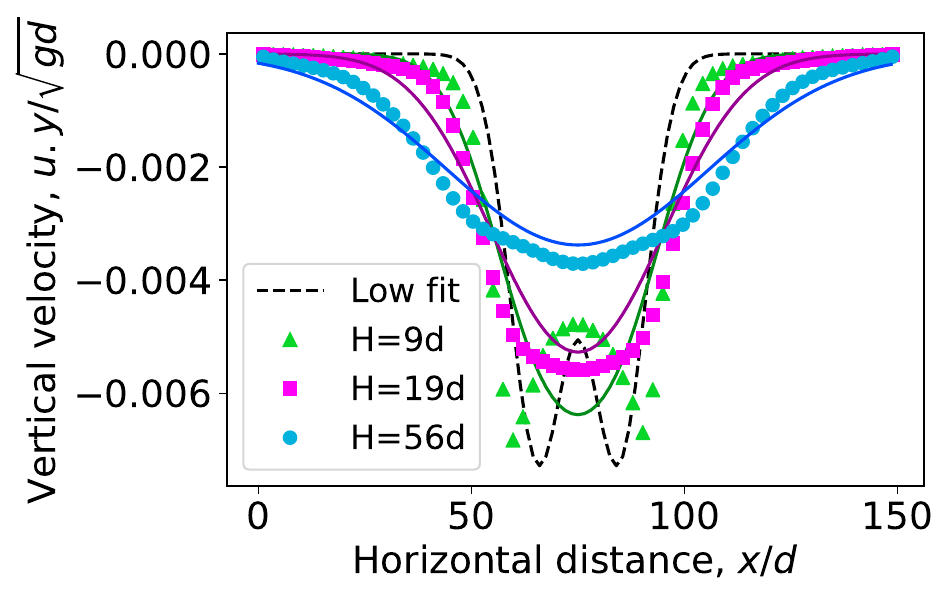}
\caption{Kinematic model (lines) compared with $\mu(I)$ model (points) in a $2D$ silo with one two openings. The vertical velocity from simulations and kinematic model are compared for various different heights. One set of kinematic parameters is fitted over the entire silo, however a set of kinematic parameters is fitted just on the $H=0.31$ case to display the double dip behaviour (indicated by the dotted line).}
\label{fig:kinematic_compare_2hole}
\end{figure}

A comparison of the behaviour for kinematic and $\mu(I)$ models in the two opening silo is given in Figure~\ref{fig:kinematic_compare_2hole}. When applying the $\mu(I)$ model to the two opening silo we obtain two dips, while the kinematic model shows almost no sign of this. The kinematic model can capture the double dip pattern if parameter values distinct from the values for other heights in the silo are used (which may improve the kinematic model for a single silo as well~\cite{choi2005velocity}). The black dotted line in Figure~\ref{fig:kinematic_compare_2hole} shows the prediction of the kinematic model using parameters fitted only over a horizontal line near the bottom of the silo. In this case the kinematic model does show the double dip behaviour we expect from this domain, although it predicts the dip occurs less gradually than the $\mu(I)$ simulations. This is possibly related to the lack of prediction for plug flow discussed in Section~\ref{sec:single_results}.

For a silo with two openings, when different separations between the openings are examined, unexpected phenomena arise. Fullard et al.~\cite{fullard2019dynamics} has shown that a `dip' occurs in mass flow rate for small separations. For silos with two orifices approximately $7.5$ particle diameters apart, a minimum flow rate is reached, where either increasing or decreasing the separation distance would increase the flow rate. It is intuitive that the flow is less than the zero separation case (which corresponds to a single opening with double width), however the observation that the flow rate for small separations is less than the flow rate for large separations is less intuitive. The Beverloo relation can be applied to the zero separation and infinite separation cases, giving the flow rates as $Q=C(2W-k)^{1.5}$ for $L=0$ and $Q=2C(W-k)^{1.5}$ for $L>>W$. If the $k$ term is omitted the flow rate in the infinite case is less than the zero separation case by a factor of $\frac{1}{\sqrt{2}}$ (approximately $0.7$), which corresponds to the value  we expect the flow rate to reach at infinite separation (with the $k$ term decreasing the limit value when accounted for). While the Beverloo relation does give the behaviour of the zero separation case and the expected infinite separation value, it does not provide any insight to the behaviour between these two cases.

\begin{figure}
\begin{tabular}{cc}
\includegraphics[width=0.4\textwidth]{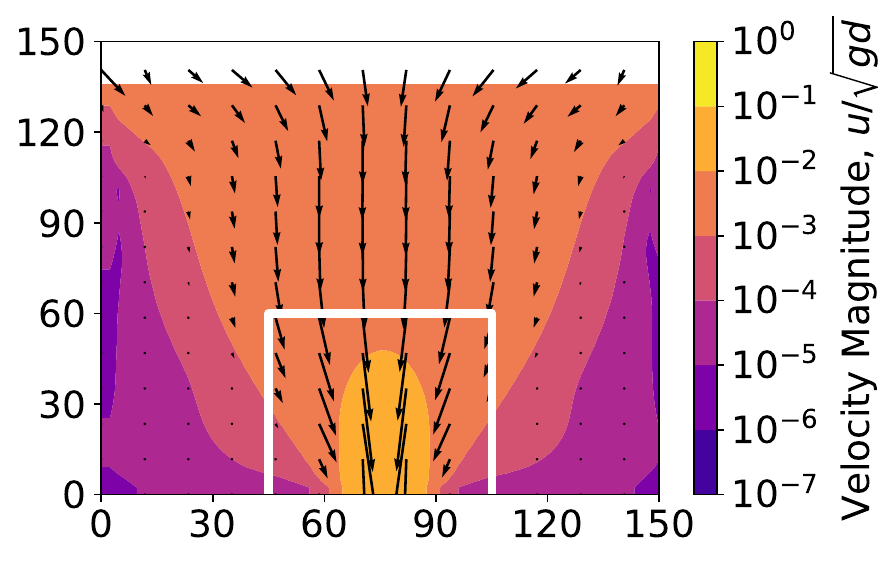} &   \includegraphics[width=0.4\textwidth]{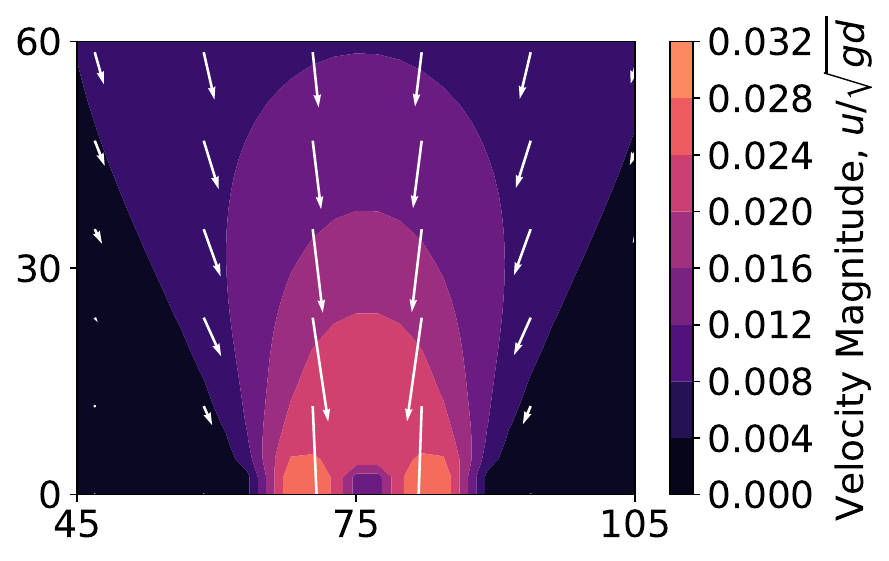} \\
Log velocity with $L/d=4.7$ & Linear velocity with $L/d=4.7$\\[6pt]
\includegraphics[width=0.4\textwidth]{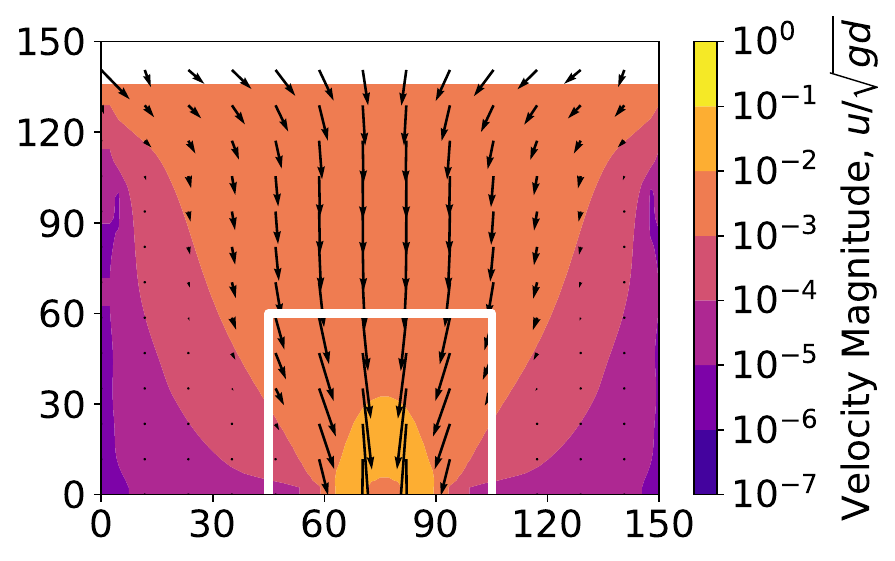} &   \includegraphics[width=0.4\textwidth]{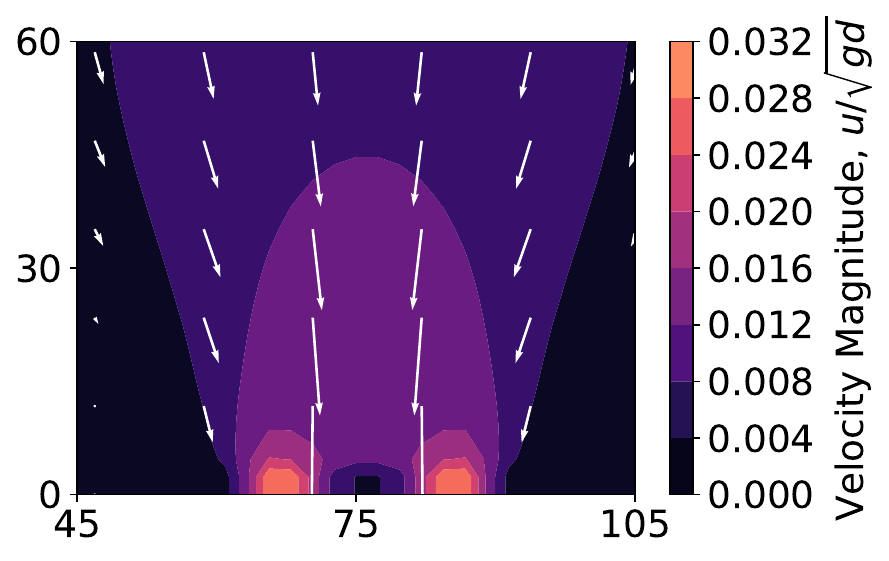} \\
Log velocity with $L/d=9.4$ & Linear velocity with $L/d=9.4$\\[6pt]
\includegraphics[width=0.4\textwidth]{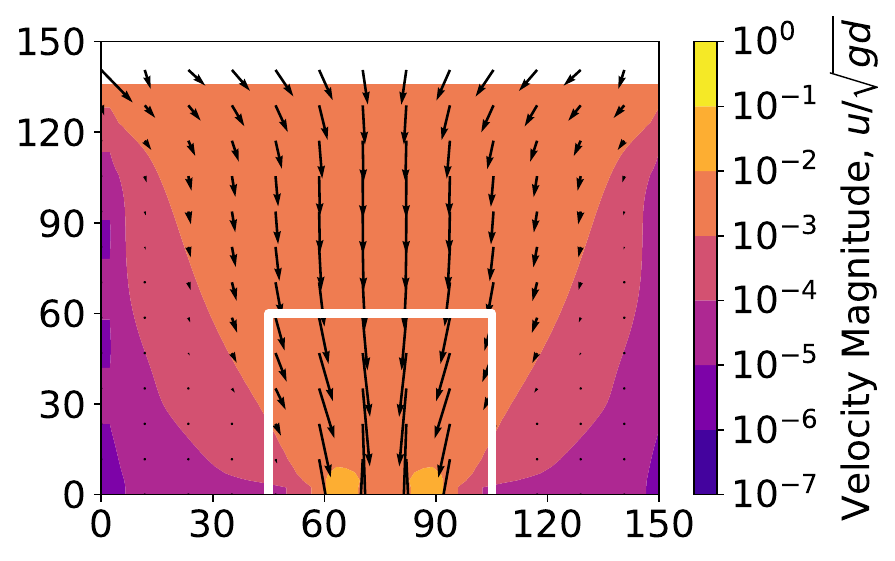} &   \includegraphics[width=0.4\textwidth]{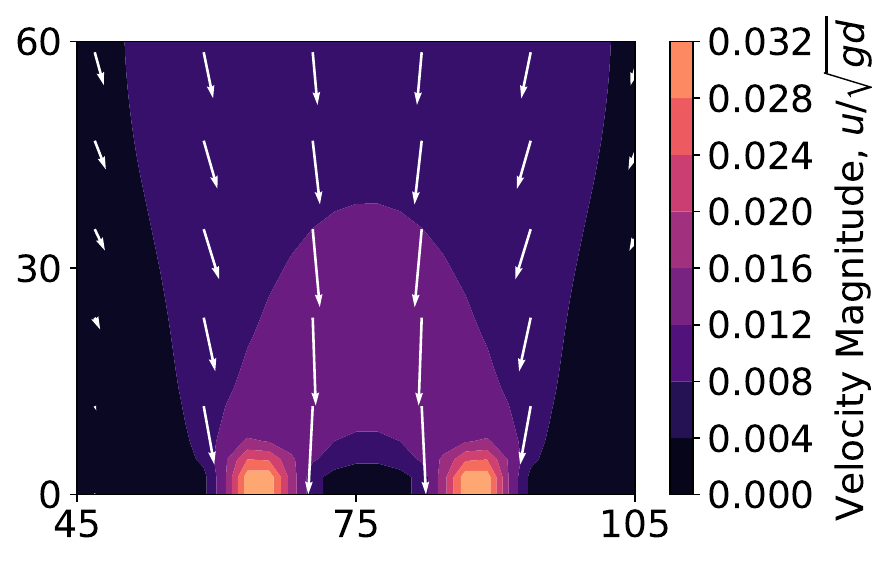} \\
Log velocity with $L/d=14.0$ & Linear velocity with $L/d=14.0$\\[6pt]
\includegraphics[width=0.4\textwidth]{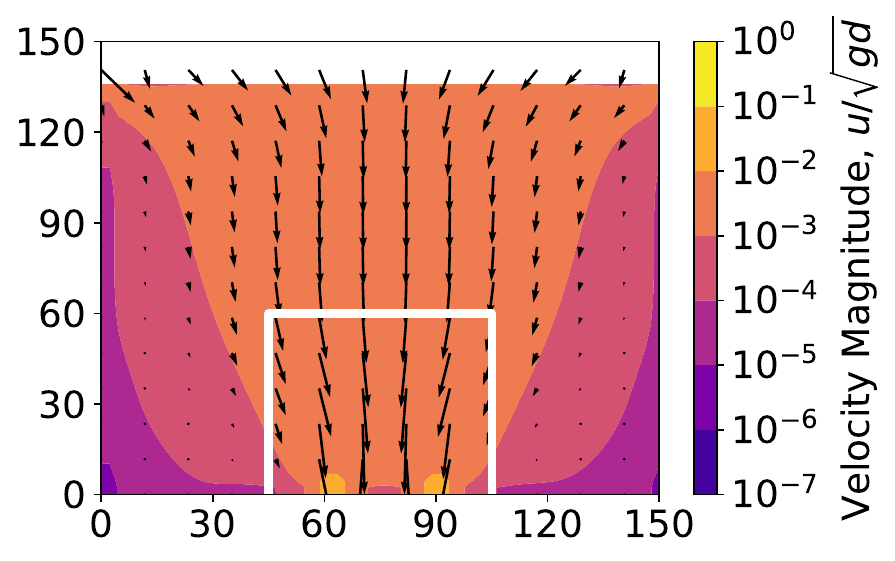} &   \includegraphics[width=0.4\textwidth]{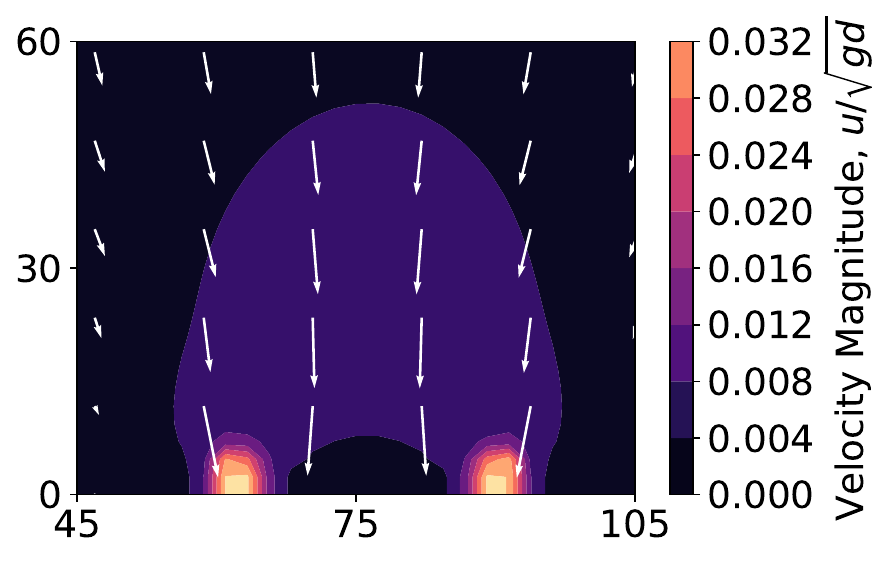} \\
Log velocity with $L/d=18.8$ & Linear velocity with $L/d=18.8$\\[6pt]
\end{tabular}
\caption{Velocity in two hole silos with various different orifice separations. The white box in the log plots indicate the area examined in the linear plots.}
\label{fig:double_contour}
\end{figure}

When examining the two opening silo over multiple separation distances, we find the $\mu(I)$ parameters (given in Equation~\ref{eq:muI}) have a big impact on the shape or existence of the flow rate dip. We examine $3$ different sets of parameters, with ($\mu_s$, $\Delta\mu$, $I_0$) being set to ($0.47$, $0.38$, $0.6$) for a low friction simulation, ($0.62$, $0.48$, $0.6$) for a medium friction simulation, and ($0.77$, $0.58$, $0.6$) for a high friction simulation. The velocity profile of two opening silos using the medium friction values for various different orifice separations is given in Figure~\ref{fig:double_contour}, with low and high friction both giving similar qualitative behaviour. For low separation distances, the fast regions near the orifice merge together creating a single fast region, while for large separations the fast regions stay relatively independent, and maintain higher independent speed.

\begin{figure}
     \centering
      \begin{subfigure}[t]{0.47\textwidth}
         \centering
         \includegraphics[width=\textwidth]{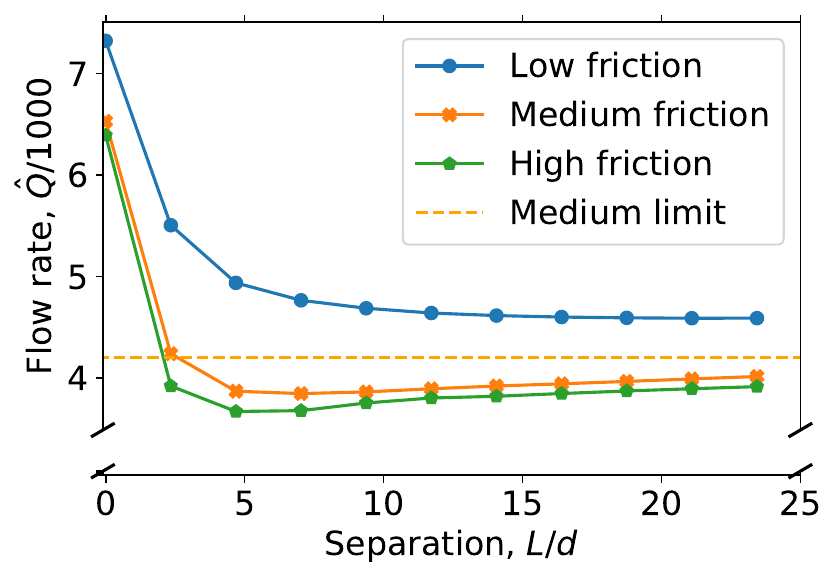}
     \end{subfigure}
     \hfill
     \begin{subfigure}[t]{0.47\textwidth}
         \centering
         \includegraphics[width=\textwidth]{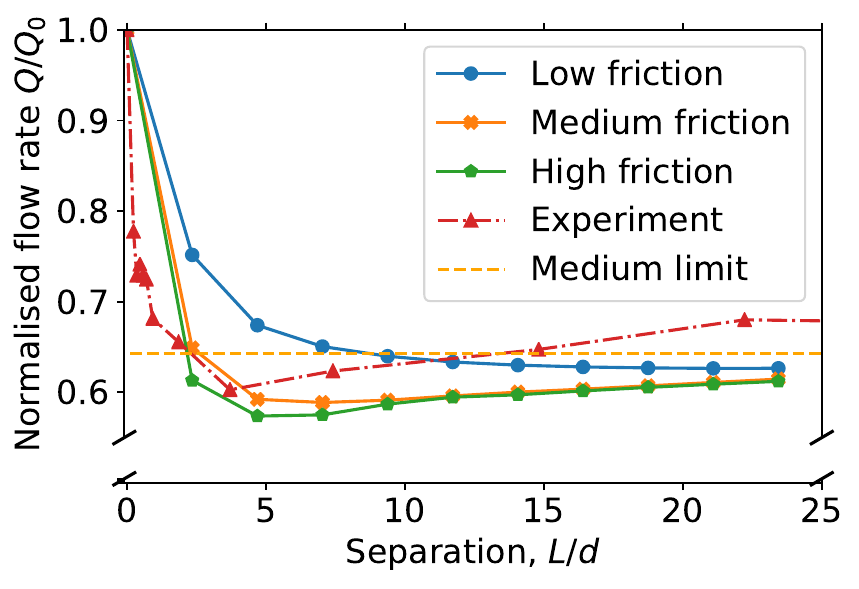}
     \end{subfigure}
        \caption{Flow rate $\hat{Q}=Q/\sqrt{gd^5}$ versus opening separation length for two opening silos with different friction values. The left plot shows the raw flow rate, while the right plot shows flow rate normalised by the zero separation case. The $\mu(I)$ parameters ($\mu_s$, $\Delta\mu$) are ($0.47$, $0.38$) for low friction, ($0.62$, $0.48$) for medium friction, and ($0.77$, $0.58$) for high friction. The doubled flow rate for a single orifice silo with medium friction values is provided for comparison (medium limit dotted line), as well as the experimental ``Amaranth small'' data from~\cite{fullard2019dynamics}}
        \label{fig:varied}
\end{figure}

The flow rates over different separations are shown in Figure~\ref{fig:varied} for different friction values and some experimental data taken from~\cite{fullard2019dynamics}. The dotted line also shows the doubled flow rate for a single opening case with the orifice width used for the double opening cases, which should give the flow rate for arbitrarily large separations (since interactions between the orifices are negligible, the silo will act as two single orifice silos). For the low friction case, we see that the flow does not dip significantly, instead decreasing in what seems to be a monotonic pattern. The low friction case does go below the expected infinite separation value; this could be because of the side walls affecting the flow for large orifice widths. For medium friction values, the flow rate dips in a similar manner to the experiment, though the dip is more gradual than the experimental data, with the dip occurring over a greater separation range than the experiment shows. The medium friction data also does not seem to have as large of a difference between the minimum dip value and the large separation value. The high friction values seem to have similar dipping behaviour to the medium friction values. Some tests with high friction values are numerically unstable, and as such for the remainder of this paper we use the `medium' values to test the other additions to the model.

\subsubsection{Hele-Shaw Wall friction}

\begin{figure}
     \centering
      \begin{subfigure}[t]{0.47\textwidth}
         \centering
         \includegraphics[width=\textwidth]{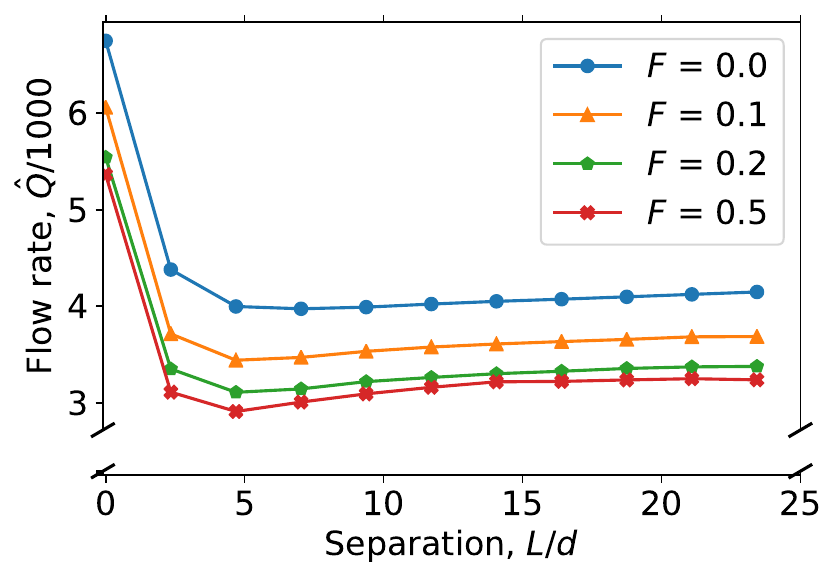}
     \end{subfigure}
     \hfill
     \begin{subfigure}[t]{0.47\textwidth}
         \centering
         \includegraphics[width=\textwidth]{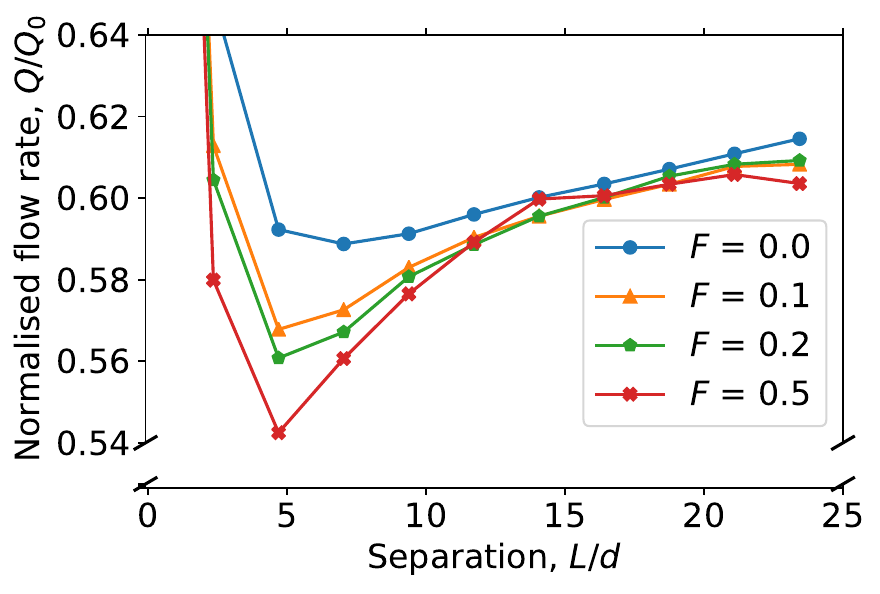}
     \end{subfigure}
        \caption{Flow rate $\hat{Q}=Q/\sqrt{gd^5}$ versus opening separation length for two opening silos with different wall friction values. The raw flow rate dip is shown on the left, while on the right the flow rate normalised by the zero separation case is shown.}
        \label{fig:wall_friction}
\end{figure}

In Figure~\ref{fig:wall_friction} we show the flow rate as a function of orifice separation for medium $\mu(I)$ parameters combined with various values of the wall friction parameter $F$. Accounting for the Hele-Shaw wall friction given by Equation~\ref{eq:hele-shaw} has a significant effect. As intuitively expected, increased friction decreases the overall flow rate for all separation values. The dipping behaviour of the flow rate occurs for all of these different wall friction values, however greater wall friction results in the dip deepening. For the normalised plot the flow rate seems to recover to the same normalised point independently of the friction values, indicating that wall friction is impactful when the separation between orifices is low while not changing the relationship between the zero separation case and the infinite separation case.

\subsubsection{Dilatancy}

\begin{figure}
     \centering
      \begin{subfigure}[t]{0.47\textwidth}
         \centering
         \includegraphics[width=\textwidth]{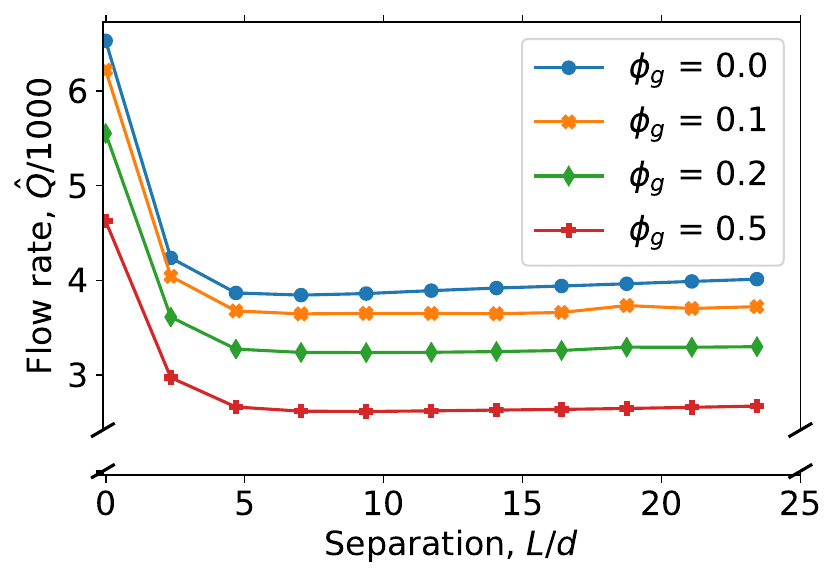}
     \end{subfigure}
     \hfill
     \begin{subfigure}[t]{0.47\textwidth}
         \centering
         \includegraphics[width=\textwidth]{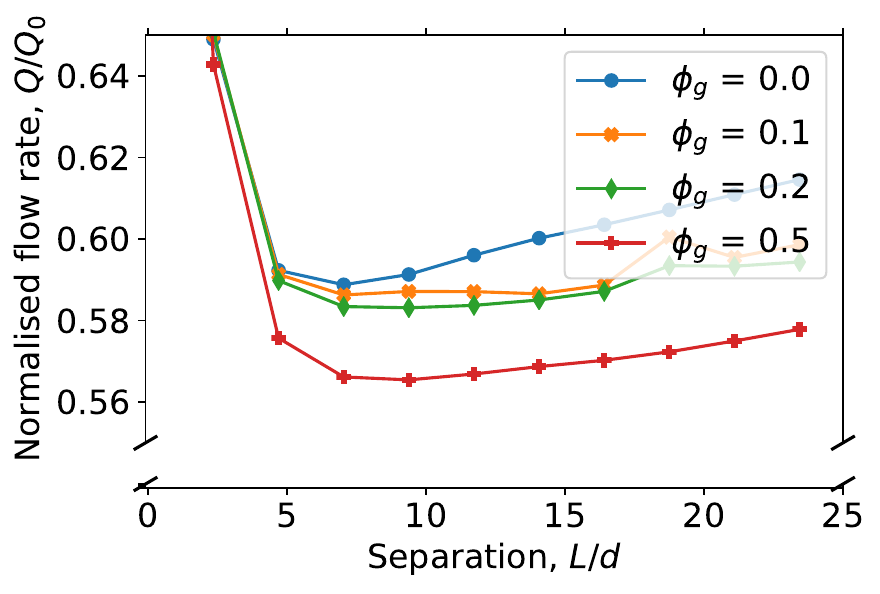}
     \end{subfigure}
        \caption{Flow rate $\hat{Q}=Q/\sqrt{gd^5}$ versus opening separation length for two opening silos with different strengths of dilatancy, determined by the value $\phi_g$. The raw flow rate dip is shown on the left, while on the right the flow rate normalised by the zero separation case is shown.}
        \label{fig:dilatancy}
\end{figure}

The effect of dilatancy on the relationship between mass flow rate and orifice separation distance is shown in Figure~\ref{fig:dilatancy}. Increasing the dilatancy decreases the mass flow rate for all separations as expected, with a realistic parameter of $0.2$~\cite{andreotti2013granular} giving approximately $20\%$ less flow than the incompressible case. The flow rate dip occurs for all parameters tested, with the normalised flow rate recovering with increased separation at similar rates. With increased dilatancy, the normalised flow dips deeper, indicating that dilatancy heavily depends on the orifice size. Since the flow rate dip recovery is somewhat consistent for the different parameters tested, dilatancy seems to primarily affect the flow rate dip by increasing the difference between the zero separation case and the non-zero separation cases.

\subsubsection{Non-local effects}

\begin{figure}
     \centering
      \begin{subfigure}[t]{0.47\textwidth}
         \centering
         \includegraphics[width=\textwidth]{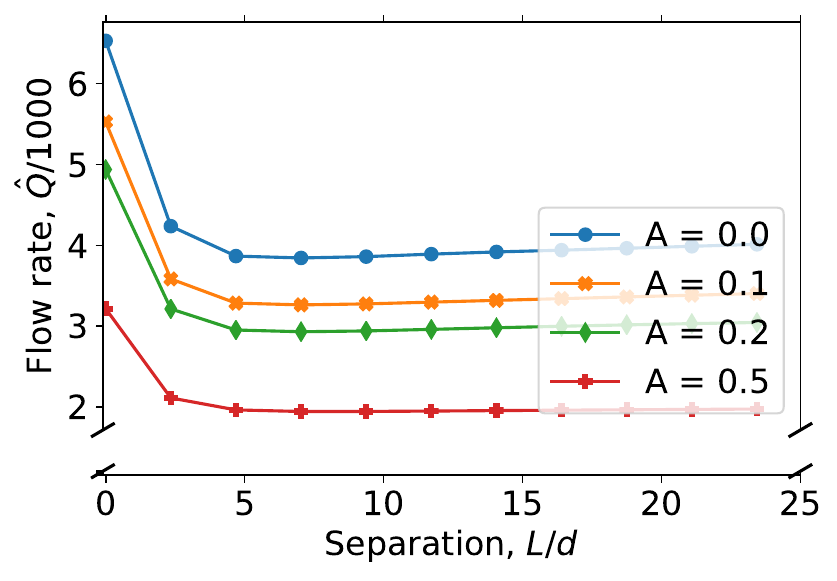}
     \end{subfigure}
     \hfill
     \begin{subfigure}[t]{0.47\textwidth}
         \centering
         \includegraphics[width=\textwidth]{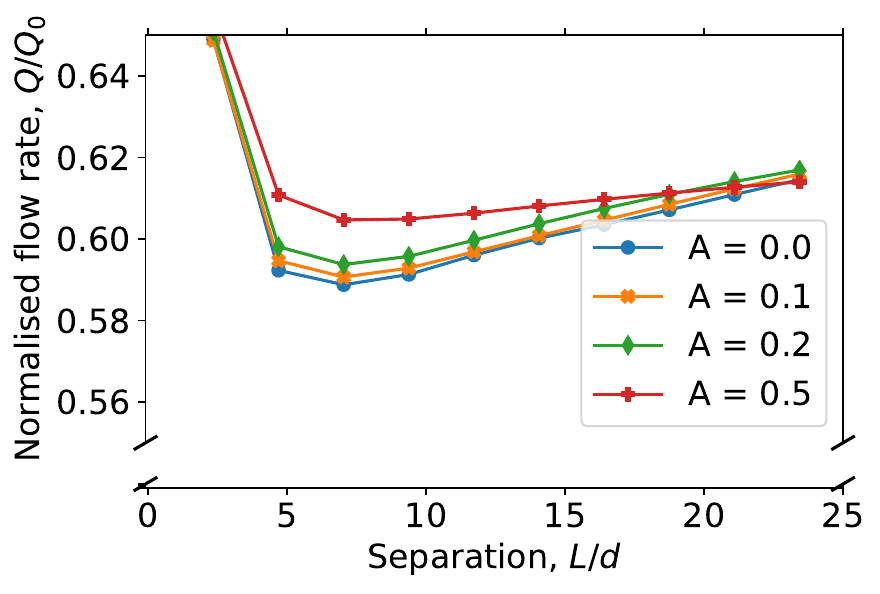}
     \end{subfigure}
        \caption{Flow rate $\hat{Q}=Q/\sqrt{gd^5}$ versus opening separation length for two opening silos with different strengths of non-local effects, as determined by parameter $A$. The raw flow rate dip is shown on the left, while on the right the flow rate normalised by the zero separation case is shown.}
        \label{fig:non_local}
\end{figure}

In Figure~\ref{fig:non_local}, non-local effects have been implemented for a two opening silo. The non-local effects are controlled by a single chosen parameter $A$. Increased non-local strength decreases the flow rate greatly, with a feasible non-local parameter $0.5$~\cite{henann2013predictive} resulting in approximately half of the flow rate compared to the local case. When examining the normalised plot, increased non-local strength seems to counteract the flow rate dip. The dip occurs for all parameters tested, however the dip is shallower when the non-local parameter $A$ is greater. The normalised flow rate seems to recover to the same value, which indicates that non-local effects has a similar impact for the single opening with double width silo and the silo with two openings separated by a large distance. However, the non-local effects do have a large impact on the flow rate dip when the distance between openings when the separation is low (but not zero). This indicates that non-local effects are important for capturing the flow rate dip.

\begin{figure}
    \centering
    \includegraphics[width=0.47\textwidth]{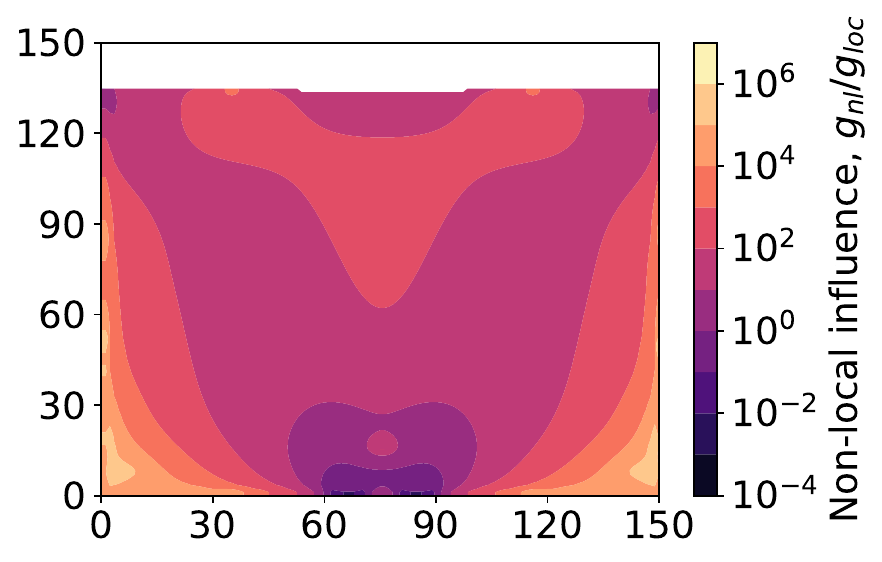}
    \caption{A log-scale contour plot of a two opening silo displaying the non-local influence $g_{nl}/g_\textrm{loc}$, which is used to show where non-local effects are strong compared to local effects.}
    \label{fig:log_xi}
\end{figure}

Figure~\ref{fig:log_xi} shows the strength of non-local fluidity $g$ normalised by the local fluidity $g_\textrm{loc}$ over a silo, which we use to represent the influence of non-local effects. This Laplacian coefficient is relatively large over most of the domain, however it drops rapidly near the orifices. This behaviour is consistent with what we expect from Equation~\ref{eq:non_local} because near the orifices $I$ is large, hence $\mu(I)$ approaches $\mu_s + \Delta\mu$, meaning that $\xi(\mu(I))$ approaches zero. This means that non-local effects are dominant over most of the silo, however near the openings non-local effects are negligible. Interestingly, in between the orifices there is a small zone for which $g/g\textrm{loc}$ is larger, indicating an increased relevance for non-local effects. This zone may explain why non-local effects decrease the impact of the flow rate dip for small separation values, since this zone may smooth out the flow for small separations.

\section{Conclusion}

We have implemented a continuum model based on the $\mu(I)$ rheology, and applied it to a two opening silo, which is a domain which displays many challenging phenomena. The $\mu(I)$ model follows the expected behaviour described by the Beverloo-Hagen relation for a single opening silo, despite not modelling single particle interactions.

For a double opening silo, the base $\mu(I)$ model is capable of capturing the transition from monotonically decreasing relationship between flow rate and orifice separation distance for low friction materials to the more complex flow rate dip which is seen in more realistic higher friction materials via the $\mu(I)$ parameters. This shows that the flow rate dip is a frictional phenomena, with higher friction being required to obtain the flow rate dip.

We also extended the $\mu(I)$ model by accounting for wall friction, dilatancy, and non-local effects. These effects all have a strong impact on the mass flow rate in each geometry we tested. We find that wall friction effects may be a significant factor for the $-kd$ shift term in the Beverloo-Hagen relation, although caution is necessary as this is partially based on theoretical data where physical silos would jam. Each of these effects also had unique impacts on the flow rate dip, with wall friction strengthening the dip, non-local effects weakening the dip, and dilatancy decreasing the flow rate for the smaller orifice sizes used for non-zero separations. The extended model that incorporates all three of these effects matches the experimental model better than the base $\mu(I)$ model.

\section{Acknowledgements}
This research was supported by the Royal Society of New Zealand (MAU1712) and by Massey University

\bibliographystyle{unsrt} 
\bibliography{aa_ref}
\end{document}